\documentclass[paper]{agujournal2019}
\usepackage{url}
\usepackage{lineno}
\usepackage{color}
\usepackage{booktabs}
\usepackage{amsmath}
\usepackage{amssymb}
\usepackage{graphicx}
\usepackage[export]{adjustbox}
\usepackage{diagbox}
\usepackage{multirow}
\draftfalse

\newcommand{\planss} {{Planatary Space Science }}  
\newcommand{\ssr}{   {Space Sci. Rev. }}

\newcommand{\jgr}{   {J. Geophys. Res.}}
\newcommand{\grl}{   {Geophys. Res. Lett.}}

\newcommand{\apjl}{   {Astrophys. J. Lett.}}

\newcommand{\aap}{   {Astronomy \& Astrophysics}}

\newcommand{\blue}{\textcolor{black}}

\journalname{JGR: Space Physics}

\begin{document}


\title{Exploring the Magnetotail from Low Altitudes: Evolution of Energetic Electron Flux During the Substorm Growth Phase}

\authors{Weiqin Sun\affil{1}, Xiao-Jia Zhang \affil{1,2}, Anton V. Artemyev\affil{2}, Rumi Nakamura\affil{3}, Jian Yang\affil{4},  Vassilis Angelopoulos\affil{2}}
\affiliation{1}{Department of Physics, University of Texas at Dallas, Richardson, Texas, USA}
\affiliation{2}{Department of Earth, Planetary, and Space Sciences, University of California, Los Angeles, Los Angeles, California, USA}
\affiliation{3}{Space Research Institute, Austrian Academy of Sciences, Graz, Austria}
\affiliation{4}{Department of Earth and Space Sciences, Southern University of Science and Technology, Shenzhen, China}

\correspondingauthor{Sun Weiqin}{weiqin.sun@utdallas.edu}

\begin{keypoints}
\item We present Rice Convection Model (RCM) simulations of magnetotail reconfiguration during the substorm growth phase
\item The RCM simulations are merged with low-altitude ELFIN observations of energetic electron fluxes
\item The RCM/ELFIN comparison confirms the interpretation of the shrinking latitudinal range of the plasma sheet projection to low altitudes
\end{keypoints}

\begin{abstract}
The magnetospheric substorm, which plays a crucial role in flux and energy transport across Earth's magnetosphere, features the formation of a thin, elongated current sheet in the magnetotail during its growth phase. This phase is characterized by a decrease in the equatorial magnetic field $B_z$ and the stretching of magnetic field lines. Observing these large-scale magnetic field reconfigurations is challenging with single-point satellite measurements, which provides only spatially-localized snapshots of system dynamics. Conversely, low-altitude spacecraft measurements of energetic electron fluxes, such as those from ELFIN, offer a unique opportunity to remotely sense the equatorial magnetic field in the magnetotail during substorms by measuring the latitudinal variations of energetic electron isotropic fluxes. Because of strong scattering caused by the curvature of magnetic field lines, energetic electrons in the magnetotail are mostly isotropic. Consequently, variations in their fluxes at low altitudes are expected to reflect the reconfiguration of the magnetotail magnetic field. 
To better understand the connection of electron flux variation at low altitudes and magnetic field reconfiguration during substorms, we compared low-altitude ELFIN observations with simulations from the Rice Convection Model (RCM). The RCM, which assumes fully isotropic electron distributions, provides a robust framework for describing energetic electron dynamics in the plasma sheet and determining the self-consistent magnetic field configuration during substorms. 
The comparison of ELFIN observations and RCM simulations confirms our interpretation of electron flux dynamics at low altitudes during the substorm growth phase and validates the use of such observations to infer magnetotail dynamics during substorms.
\end{abstract}

\section{Introduction}
The magnetospheric substorm, a highly energetic and key phenomenon in Earth's magnetosphere, begins with a growth phase characterized by the formation of a thin, elongated current sheet in the magnetotail \cite<known as current sheet thinning, see, e.g.,>{Angelopoulos08,Baker96,Sitnov19,Runov21:jastp}. This thinning is crucial for the flux transport and energy conversion process during substorms \cite<e.g.,>{Sergeev93}. Instabilities in these thin current sheets are believed to drive magnetic field line reconnection and current disruption, resulting in a complex magnetic field reconfiguration during the substorm expansion phase \cite<see>[and references therein]{Sitnov19}. Therefore, a thorough study of the thinning and structural evolution of the current sheet during the growth phase is important to fully understand the intricate particle and energy dynamics within the magnetotail, both before and throughout the progression of a substorm.\par

The current sheet thinning during the substorm growth phase has been well documented through observations from near-equatorial spacecraft \cite<e.g.,>{Artemyev16:jgr:thinning,Petrukovich07,Sergeev11,Yushkov21} and numerical simulations \cite<e.g.,>{Birn04MHD,Gordeev17,Lu19:jgr:cs,Lu18:3dthinning}. Key signatures of current sheet thinning include a decrease in the equatorial magnetic field $B_{z}$ (indicative of magnetic field line stretching), an increase in equatorial current density $j_{y}$, and an elevation in the lobe magnetic field $B_{L}$ \cite<e.g.,>{Birn98:cs,Schindler&Birn82,Schindler&Birn93}. Despite these advancements, the spatial extent of current sheet reconfiguration remains poorly understood. Near-equatorial missions, with orbital periods of tens of hours, can effectively capture the temporal variability and reconfiguration of the current sheet during substorms. However, they provide limited insights into the physical scale of the thin current sheet, which may span  the entire near-Earth magnetotail \cite<e.g.,>{Artemyev15:grl:2dCS,Gordeev17, Hsieh&Otto15,Sitnov19:jgr} or even extend from the near-Earth region to as far as the lunar orbit \cite<e.g.,>{Angelopoulos13, Artemyev19:jgr:globalview}. A powerful alternative for probing the magnetotail current sheet configuration across a wide radial ($X$) distance is remote sensing via low-altitude, polar-orbiting spacecraft \cite<e.g.,>{Wing&Newell02,Sergeev23:elfin}. These spacecraft traverse the low-altitude projection of the entire near-Earth magnetotail in minutes, providing a snapshot-like view of equatorial electron flux dynamics with high spatial resolution in the radial direction. Recent high-resolution electron measurements from the low-altitude Electron Losses and Fields Investigation (ELFIN) CubeSats \cite{Angelopoulos20:elfin} have demonstrated that the equatorial current sheet thinning can be effectively monitored by observing the dynamical evolution of plasma sheet (PS) energetic electron fluxes at low altitudes during the substorm growth phase. \blue{These observations support the hypothesis that the shrinking of the latitudinal extent of the PS magnetic projection—defined here as the low-altitude ionospheric footprint of the near-equatorial plasma sheet characterized by isotropic energetic electron fluxes in the 50–300 keV range—is a consequence of magnetotail magnetic field reconfiguration during the substorm growth phase \cite{Artemyev22:jgr:ELFIN&THEMIS}.}

To confirm and refine the interpretation of low-altitude PS electron flux observations during \blue{substorm growth phase}, we combine numerical simulations from the Rice Convection Model (RCM) with measurements from ELFIN’s statistical observations \cite{Artemyev22:jgr:ELFIN&THEMIS}. We compare the energetic electron fluxes and energy spectra of RCM and ELFIN during current sheet thinning. The RCM is a well-established, first-principles model of Earth's magnetosphere \cite<>[and references therein]{Toffoletto03}. It calculates both \( E \times B \) and gradient/curvature drift velocities \( v_k = ((E - \lambda_k \nabla V^{-2/3}) \times B)/B^2 \) for isotropically distributed, \(\lambda\)-conserving particles \cite{Wolf83}. Here \( E\) and \(B\) represent the electric and magnetic fields, respectively, and \(\lambda_k=W_kV^{2/3}\) is the energy invariant, conserved as particles drift within flux tubes filled with isotropic particles of kinetic energy \(W_k\) \cite{Wolf83,Schulz&Chen08}. The quantity \(V=\int ds/B\) is the flux tube volume per unit magnetic flux, which extends across the magnetosphere. The particle distribution function \( f_k \) follows the advection equation \( \left( \partial/\partial t + v_k \cdot \nabla \right) f_k = S(f_k) - L(f_k) \) \cite{Wolf83}, where \( S(f_k) \) and \( L(f_k) \) are sources and sinks, respectively. The subscript \( k \) represents particles with specific charge, mass, and energy invariants \(\lambda_k\). The electric \( E \) and magnetic field \( B \) are self-consistently calculated \cite{Toffoletto03,Yang19}. The electric potential field is determined by solving the current conservation equation that couples the currents between the magnetosphere and the ionosphere. The magnetic field profiles are self-consistently calculated using a finite-volume MHD code integrated with the RCM \cite{Silin13,Yang19}. This self-consistent framework makes the RCM particularly well-suited for tracing energetic electrons in the magnetotail, where these particles are usually isotropic \cite{Artemyev22:jgr:ELFIN&THEMIS} and follow both \( E \times B \) and gradient/curvature drifts \cite<e.g.,>{Birn14,Gabrielse19}.  Although the RCM's isotropic assumption limits its ability to determine the plasma sheet's inner boundary via the precipitating-to-trapped flux ratio, the ratio of magnetic field line curvature radius (\(R_c\)) to particle gyroradius (\(\rho\)) provides a useful criterion. Specifically, this ratio predicts nonadiabatic scattering when \(R_c/\rho < 8\) and loss cone depletion when \(R_c/\rho > 8\) \cite{Sergeev&Tsyganenko82,Delcourt94:scattering}. Thus, RCM simulations offer valuable insights into electron dynamics during substorm growth phases, complementing low-altitude observations.

This study aims to combine RCM simulations, which reproduce magnetic field reconfiguration during the substorm growth phase, with measurements of energetic electron fluxes by ELFIN prior to substorms. By modeling the magnetotail reconfiguration and comparing the simulated evolution of low-altitude electron fluxes with ELFIN observations during the growth phase, we seek to confirm the critical role of magnetic field reconfiguration in driving low-altitude energetic electron dynamics. The paper is organized in five sections: Section~\ref{sec:data} provides a detailed description of the ELFIN instruments used for energetic electron measurements and highlights representative examples of ELFIN observations during the substorm growth phase. Section~\ref{sec:model} shows the RCM  simulations during the substorm growth phase, based on ELFIN observations from Section~\ref{sec:data}. In Section~\ref{sec:comparison}, we present a thorough comparison between the RCM simulation results and ELFIN observations, emphasizing key similarities and discrepancies. Finally, Section~\ref{sec:discussion} discusses the broader implications of our findings, interprets the results in the context of substorm dynamics, and summarizes the study's contributions to advancing our understanding of these processes.

\section{ELFIN Observations}\label{sec:data}
ELFIN consists of two identical CubeSats equipped with energetic particle detectors (EPD) measuring ion and electron distributions within $[50,6000]$keV, at a spin resolution of $3$ seconds \cite{Angelopoulos20:elfin}. Due to ELFIN's spinning, the EPD covers the entire pitch-angle range $[0,360^\circ]$ every $3$ seconds with an angular resolution of $22.5^\circ$. In this study, we use two data products of electron EPD: the spectrum of precipitating electrons (within the local bounce loss cone), denoted as $J_{loss}$, and the spectrum of locally trapped electrons (outside the local bounce loss cone), denoted as $J_{trap}$ \cite{Angelopoulos23:ssr,Tsai24:review}. The ratio $J_{loss}/J_{trap}$ provides insights into electron scattering near the equatorial plane: an isotropic ratio of $J_{loss}/J_{trap}\sim 1$ indicates strong diffusion \cite{Kennel69}, which is either associated with electron scattering by magnetic field line curvature in the plasma sheet \cite{Wilkins23, Artemyev23:ELFIN&dispersion} or wave-particle resonant scattering in the inner magnetosphere \cite<e.g.,>{Tsai22,Chen22:microbursts,Zhang22:microbursts,Grach22:elfin}. The boundary between the plasma sheet \cite<characterized by isotropic electron fluxes with energies up to around 100-300 keV; see>{Artemyev22:jgr:ELFIN&THEMIS} and the inner magnetosphere \cite<a region with anisotropic electron fluxes where $J_{loss}/J_{trap}\sim 0$ and sporadic isotropic precipitation due to wave-particle interactions; see>[and references therein]{Zhang23:jgr:ELFIN&scales} is known as the isotropy boundary (IB). This boundary is defined by the transition from $J_{loss}/J_{trap}\sim 1$ at higher latitudes to $J_{loss}/J_{trap}\sim 0$ at lower latitudes \cite<see details in>{Sergeev12:IB,Sergeev23:elfin}. \blue{For a fixed electron energy, the isotropy boundary can be determined from the precipitating-to-trapped flux ratio plotted along the magnetic latitude of the spacecraft orbit \cite<e.g.,>{Sergeev12:IB}. For a wide energy range, however, such single-energy latitudes of the isotropy boundary form some curve that can be called an isotropy boundary pattern \cite<see discussion in>{Artemyev24:jgr:ELFIN&IBe}. For simple monotonic isotropy boundary patterns (when magnetic latitude increases with an energy increase), this pattern can be traced by the contour line of the precipitating-to-trapped flux ratio \cite<see examples in>{Wilkins23,Artemyev24:jgr:ELFIN&IBe}, whereas non-monotonic patterns of the isotropy boundary require more sophisticated approaches \cite<see examples in>{Shi&Stephens24:arxiv}. In this study, we follow the simple approach and use contour lines of the precipitating-to-trapped flux ratio.}
The dynamics of this boundary during substorms reflect the reconfiguration of the magnetotail magnetic field, particularly the earthward propagation of the curvature scattering region (thin magnetotail current sheet) \cite<see details in>{Sergeev12:IB,Sergeev23:elfin,Artemyev22:jgr:ELFIN&THEMIS,Shi&Stephens24:arxiv}. To verify this interpretation of low-altitude ELFIN measurements of energetic electron fluxes, we compare several substorm events from ELFIN’s statistical observations \cite{Artemyev22:jgr:ELFIN&THEMIS} with simulations of magnetotail dynamics.

\begin{figure}[!htbp]
    \centering
    \includegraphics[width=1\linewidth]{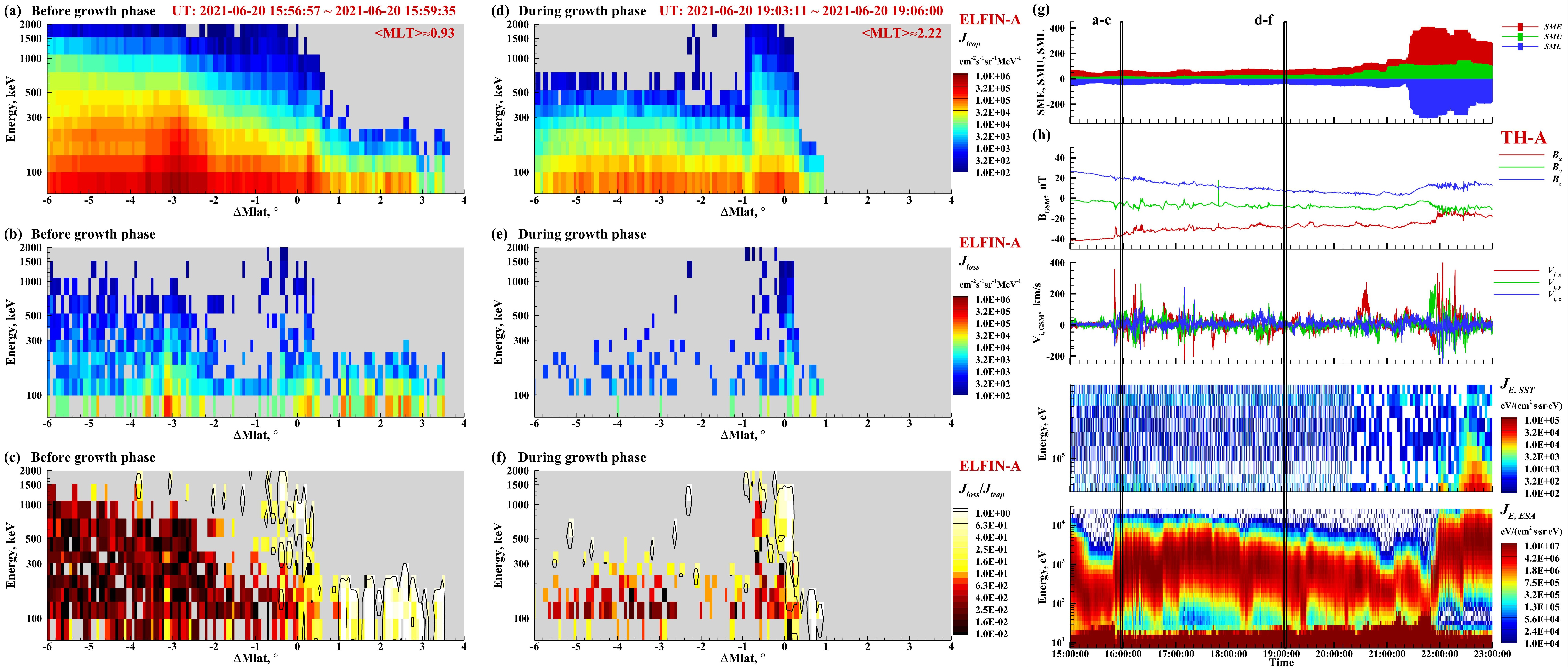}
    \caption{Overview of the first ELFIN event, showing measurements collected on two orbits: one before a substorm growth phase (a-c) and the other during the substorm growth phase (d-f). \blue{The collection times and corresponding average MLT are shown above each panel}, with the pre-growth phase one occurring 3 hours prior to the substorm growth phase. The panels display quiet-time ELFIN measurements of locally trapped fluxes (a, d), precipitating fluxes (b, e), and the precipitating-to-trapped flux ratio (c, f), respectively. \blue{Grey regions correspond to bins with insufficient counts (\(<4\)). \(Mlat_0\) is defined as the magnetic latitude at the center (approximately for $500$ keV) of the narrow interface region between the outer radiation belt and the plasma sheet.} The magnetic latitude is then shifted to \(Mlat-Mlat_0\), resulting in \(\Delta Mlat < 0\) for the outer radiation belt and \(\Delta Mlat > 0\) for the plasma sheet. \blue{The \(J_{\text{loss}}/J_{\text{trap}} \approx 0.25\) contour indicates the isotropy boundary in panels (c) and (f). Panel (g) shows SM geomagnetic indices, and panel (h) shows near-equatorial THEMIS observations (from top to bottom): magnetic field vector, plasma flow speed, energetic ($>30$keV) and plasma sheet ($<30$kev) electron fluxes.}}
    \label{fig1}
\end{figure}

Figure \ref{fig1} shows the first event, comparing ELFIN measurements during two orbits: one before the substorm (Fig. \ref{fig1}a-\ref{fig1}c) and the other during the substorm growth phase (Fig. \ref{fig1}d-\ref{fig1}f), \blue{with their time intervals and average MLTs labeled above each set of panels. The classification of these two time intervals relative to the substorm phases is based on the background observations shown in Fig. \ref{fig1}g and \ref{fig1}h, which include SM geomagnetic indices \cite<analog of AL/AU/AE indices; see>{Gjerloev12:supermag} and near-equatorial THEMIS measurements (magnetic field vector, plasma flow speed, and energetic and plasma sheet electron fluxes). These supporting data allow us to determine the timing of the substorm onset and to establish the correspondence between the ELFIN observation intervals and the substorm growth phase. The SM geomagnetic indices in Fig. \ref{fig1}g suggest that a substorm onset occurred at $\sim 21:30$UT, preceded by a notably prolonged growth phase. To verify that ELFIN at $\sim 19:00$ UT detected the thinning current sheet during substorm growth phase, we examine THEMIS near-equatorial observations from the night-side MLT sector at $9.3 \sim 13.2 R_E$  \cite{Angelopoulos08:ssr}. Fig. \ref{fig1}h shows a $B_z$ decrease \cite<measured by the flux-gate magnetometer; see>{Auster08:THEMIS} and an absence of plasma flows before the dipolarization at $\sim 21:30$UT. The spectrum of plasma sheet electrons (measured by the electrostatic analyzer for energies $<30$keV; see \citeA{McFadden08:THEMIS}) shows continuous decrease of electron energies -- very typical signature of the near-Earth current sheet thinning during the growth phase \cite<see>{Artemyev19:jgr:globalview}. Energetic electrons (measured by the solid state detector for energies $>30$keV; see \citeA{Angelopoulos08:sst}) appear only after substorm onset and dipolarization. Therefore, both THEMIS and SM indices indicate that ELFIN observations around $\sim 19:00$ UT are associated with the early stage of the substorm growth phase.}

\blue{For each orbit, ELFIN measurements show trapped fluxes (Fig. \ref{fig1}a and Fig. \ref{fig1}d), precipitating fluxes (Fig. \ref{fig1}b and Fig. \ref{fig1}e), and the precipitating-to-trapped flux ratio (Fig. \ref{fig1}c and Fig. \ref{fig1}f). The grey areas correspond to bins with insufficient counts (less than 4), where flux values are set to NaN \cite{Angelopoulos23:ssr}. We use the contour line of precipitating-to-trapped flux ratio $\approx 0.25$ to determine the boundary between isotropic and anisotropic fluxes in the (latitude, energy domain) in Fig. \ref{fig1}c and Fig. \ref{fig1}f. The specific number of the flux ratio is not important, until it reflects the transition from weak to strong precipitation. This transition (boundary) is the low-altitude projection of the equatorial magnetic field gradient, and can be quite sharp, without sufficient ELFIN spins to resolve the interval gradient of the precipitating-to-trapped ratio \cite<see different examples in>{Artemyev24:jgr:IBe}. This is exactly the case for the event from Fig. \ref{fig1}, where additional fragmentation of precipitating fluxes further complicates the determination of latitudinal gradient of the precipitating-to-trapped ratio. Thus, we use quite moderate value ($0.25$) for this boundary determination. Fig. \ref{fig1}c and Fig. \ref{fig1}f show that $\Delta Mlat\approx 0$ when the precipitating-to-trapped flux ratio crosses the $0.25$ value within $[300,500]$keV energy range.} 

\blue{For each set of panels (Figs. \ref{fig1}a-\ref{fig1}c and Figs. \ref{fig1}d-\ref{fig1}f), the latitude \(Mlat_0\) is defined as the midpoint of the narrow interface region between the outer radiation belt (characterized by anisotropic fluxes with \(J_{loss}/J_{trap} \sim 0\)) and the plasma sheet (where \(J_{loss}/J_{trap} \sim 1\)). Although the exact IB latitude varies slightly with energy, this range is narrow and not fully resolved by ELFIN's latitudinal resolution. Therefore, we select the center value (approximately for $500$ keV) of this narrow magnetic latitude range as a representative \(Mlat_0\) for simplicity. The latitude is then shifted to a relative coordinate system as \(\Delta Mlat = Mlat - Mlat_0\), where \(\Delta Mlat < 0\) corresponds to the outer radiation belt and \(\Delta Mlat > 0\) to the plasma sheet. This normalization allows for a direct comparison of the spatial extent of the low-altitude plasma sheet projection across different substorm phases.} These panels indicate: (1) at high latitudes (\(\Delta Mlat > 0\)), ELFIN observes the projection of the plasma sheet with isotropic fluxes below 300 keV; (2) at lower latitudes (\(\Delta Mlat < 0\)), \blue{ELFIN detects both sub-relativistic and relativistic fluxes (including those above 500 keV) with minimal precipitating flux (\(J_{loss}/J_{trap} \sim 0\))}; and (3) in the transition region around \(\Delta Mlat \sim 0\), the isotropy boundary shows a gradual increase in isotropic precipitating flux energies (\(J_{loss}/J_{trap} \sim 1\)) toward lower latitudes \cite<see>{Wilkins23}. 
The comparison between Fig. \ref{fig1}a-\ref{fig1}c and \ref{fig1}d-\ref{fig1}f demonstrates that during current sheet thinning, the plasma sheet projection (\(\Delta Mlat > 0\)) at low altitudes contracts into a narrow interface region adjacent to the outer radiation belt. This effect is primarily due to the magnetic field line projection during the substorm growth phase. Furthermore, Figure \ref{fig1}d shows that during thinning, flux magnitudes comparable to pre-thinning levels are confined within the interface region of \blue{\(|Mlat| < 1^\circ\)}, confirming that the plasma sheet shrinks into an extremely narrow interface region during the event. Note that the comparison between Fig. \ref{fig1}a-\ref{fig1}c and \ref{fig1}d-\ref{fig1}f also shows significant changes in the nominal radiation belt fluxes ($Mlat_0<0$), but these changes are primarily attributable to the difference in $MLT$ of the two ELFIN orbits. The fluxes of low-altitude trapped electrons, located just a few degrees away from the bounce loss cone and characterized by strong anisotropy within the outer radiation belt, exhibit large variation with local time \cite<see Fig. 2 in>{Mourenas21:jgr:ELFIN}, but this variation is unrelated to the dynamics of low-altitude isotropic fluxes in the plasma sheet ($Mlat_0>0$) \cite<see>[for examples of multiple ELFIN orbits during a single substorm]{Artemyev22:jgr:ELFIN&THEMIS}.

\begin{figure}[!htbp]
    \centering
    \includegraphics[width=1\linewidth]{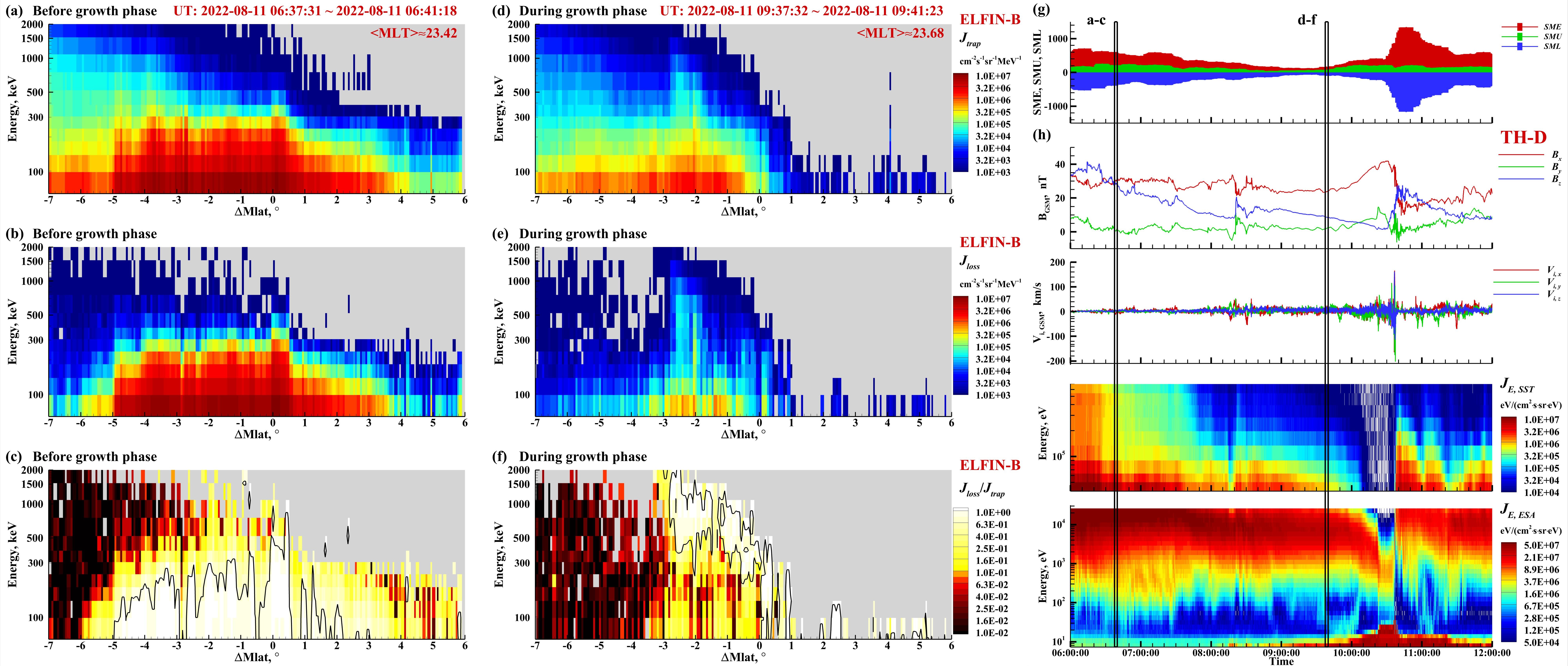}
    \caption{Overview of another ELFIN event, presented in the same format as Figure \ref{fig1}.}
    \label{fig2}
\end{figure}

\blue{Figure \ref{fig2} shows another event with the latitudinal shrinking of low-altitude projection of the plasma sheet, but with a more complex substorm background than that in Figure \ref{fig1}.} The flux magnitudes in Figure \ref{fig2} are significantly higher, about an order of magnitude greater than those in Figure \ref{fig1}, indicating stronger energization in the plasma sheet during the substorm preceding this event. \blue{Indeed, analysis of SM indices and THEMIS near-equatorial observations show much more geomagnetically disturbed conditions. Fig. \ref{fig2}g with the time profiles of SM indices show that ELFIN observations around $\sim 06:45$UT are during substorm expansion phase, whereas ELFIN observations around $\sim 09:45$UT are during the growth phase of the next substorm with an onset at $\sim10:30$UT. THEMIS observations in Fig. \ref{fig2}h support this conclusion: between plasma sheet injection at $\sim 08:20$UT and the substorm onset (with $B_z$ dipolarization and energetic electron flux enhancement) at $\sim10:30$UT, THEMIS detected the current sheet thinning characterized by $B_z$ decrease and $B_x$ increase \cite<see>{Petrukovich07,Artemyev16:jgr:thinning}.}

\blue{The enhanced fluxes (including more precipitating fluxes) in this event allow us to use a higher precipitating-to-trapped flux level, $0.75$, to draw the boundary between isotropic and anisotropic fluxes. For the first ELFIN orbit, shown in Fig. \ref{fig2}a-\ref{fig2}c, this contour level shows precipitation bursts likely generated by whistler-mode waves \cite<see detailed analysis of similar ELFIN observations in>{Tsai22,Artemyev24:jgr:ELFIN&injection,Kang24:elfin}. Thus the dynamical conditions do not allow the identification of the isotropy boundary \cite<see discussion in>{Artemyev24:jgr:ELFIN&IBe}, but ELFIN observations show the main feature for our simulations - a long plasma sheet spanning a large latitudinal range. Note that we set $\Delta Mlat=0$ around the nominal transition between the plasma sheet (fluxes are only for $<300$keV) and the outer radiation belt (relativistic fluxes), but $Mlat_0$ magnetic latitude does not precisely trace isotropy boundary, which cannot be localized for this ELFIN orbit. However, ELFIN observations at this orbit clearly show a plasma sheet extending across a wide latitudinal range and an elevated level of energetic electron fluxes. Thus, we may use these measurements (specifically, the latitudinal range of the plasma sheet projection to low altitude) as initial conditions for our simulations.}

\blue{The second ELFIN orbit, from Fig. \ref{fig2}d-\ref{fig2}f, shows a much clearer isotropy boundary separating isotropic and anisotropic electron fluxes. We set $\Delta Mlat=0$ at the midpoint of the contour level of the precipitating-to-trapped flux ratio at $\approx 0.75$ for $[300,500]$keV. Note that there are two strong precipitation bursts of relativistic electrons equatorward from $\Delta Mlat=0$, and these bursts can represent a prolongation of the isotropy boundary or may result from electron precipitation driven by electromagnetic ion cyclotron waves \cite<see analysis of similar events in>{Artemyev23:ELFIN&dispersion}.}

Comparing the panels before and during the growth phase, this event displays an even more pronounced reduction in energetic electron fluxes within the plasma sheet during the growth phase. The change in radiation belt fluxes ($Mlat_0<0$) can be attributed to variations in $MLT$ of low-altitude trapped flux measurements \cite<see Fig. 2 in>{Mourenas21:jgr:ELFIN}. Next, we will compare these two events shown in Figs. \ref{fig1}, \ref{fig2} with RCM simulations of substorm magnetotail dynamics.

\section{RCM: Substorm Growth Phase}\label{sec:model}
In this study, we utilize the Rice Convection Model (RCM), which assumes fully isotropic electron distributions, to simulate typical processes during the substorm growth phase in Earth's plasma sheet. During this phase, strong pitch-angle scattering effectively produces nearly isotropic distributions of $>1$keV electrons \cite<e.g.,>{Artemyev14:jgr}. The initial RCM-E simulations of the substorm growth and expansion phases \cite{toffo2000} introduced a force-balanced magnetic field integrated with plasma pressure derived from the RCM. Building on this foundation, further substorm simulations \cite<e.g.,>{Yang09phd} demonstrated that carefully adjusting tailward plasma boundary conditions enabled RCM to accurately reproduce energetic proton flux variations observed at geosynchronous orbit \cite{Yang08}. Subsequent studies \cite{Zhang09:rice:convection1, Zhang09:rice:convection} advanced RCM-based analyses by enhancing the Tsyganenko model with algorithms to simulate magnetic field line stretching and dipolarization. \blue{Consistent with previous RCM simulation studies \cite{Yang11, Yang13, Yang14}}, our simulations use initial and boundary conditions based on the empirical T89 magnetic field model \cite{Tsyganenko89} and plasma distribution models \cite{Lemon03, Tsyganenko&Mukai03}, driven by typical solar wind parameters and geomagnetic indices during geomagnetically quiet times \blue{(including a solar wind speed of $\sim$310 km/s, density of $\sim$14 cm\(^{-3}\), IMF \(B_z\) near -3 nT, and a geomagnetic index \(Kp \sim 1\)).} The potential drop across the polar cap is set to 40 kV, as estimated from equation (12) of \citeA{Zhang09:rice:convection}. The current thinning mechanism in the RCM reflects the energy buildup in the magnetotail during the substorm growth phase, leading to magnetic field reconfiguration. While various physical mechanisms can drive current sheet thinning \cite<e.g.,>{Sitnov19,Runov21:jastp}, the RCM focuses on capturing the ionospheric response rather than the specific origin of thinning. The goal of this study is to simulate the energetic electron fluxes observed at low altitudes during current sheet thinning, ensuring that the results align with ELFIN observations as understood from prior studies, rather than being dependent on the exact driver of thinning in the RCM. Therefore, the primary requirement for RCM is to replicate key features of substorm growth phase dynamics,  specifically magnetic field line stretching and equatorial magnetic field reduction. \par

\begin{figure}[!htbp]
    \centering
    \includegraphics[width=1\linewidth]{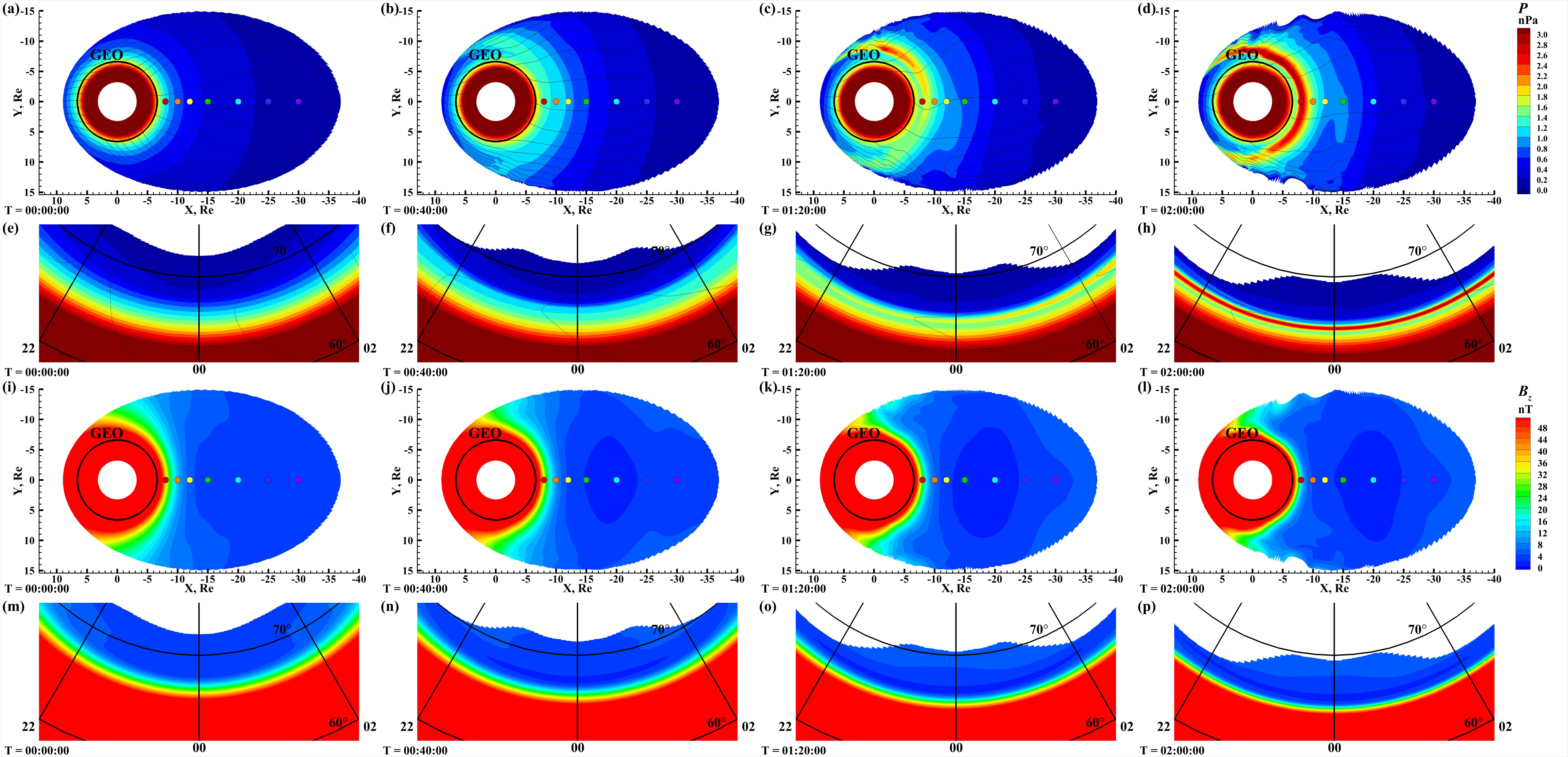}
    \caption{Overview of the RCM simulation results for the 2-hour substorm growth phase in both the RCM equatorial plane and ionosphere. (a-d) The pressure distribution in the RCM equatorial plane at various times, with colored circles marking the positions corresponding to the time series in Figure \ref{fig4}. (e-h) Nightside ionospheric pressure distribution between 22 $\sim$ 02 MLT and \(60^\circ \sim 70^\circ\). (i-p) The magnetic field \(B_z\) distribution in the RCM equatorial plane and nightside ionosphere over time.}
    \label{fig3}
\end{figure}

Figure \ref{fig3} provides an overview of the RCM simulation results for the 2-hour substorm growth phase, showing both the equatorial plane and ionosphere. The simulation incorporates self-consistent electric and magnetic fields, with a high-resolution magnetosphere-ionosphere (M-I) coupling scheme that calculates magnetic-field-aligned currents connecting the magnetosphere to the ionosphere. Fig. \ref{fig3}a-\ref{fig3}d show the pressure distribution in the RCM equatorial plane at different times. As the plasma sheet thins during the substorm growth phase,  pressure in the equatorial plane progressively increases over time, leading to the formation of a distinct high-pressure region with a sharp gradient near the geosynchronous orbit. \blue{Fig. \ref{fig3}e-\ref{fig3}h present the ionospheric perspective of the pressure variation, showing the nightside ionospheric pressure distribution over 22 $\sim$ 02 MLT and \(60^\circ \sim 70^\circ\), with thin black lines indicating contours of constant MLT and magnetic latitude.} In addition to the obvious formation of the high-pressure region, the nightside ionospheric view also reveals a significant reduction in the magnetic latitude corresponding to the RCM tail boundary near 37 $R_E$. This reduction indicates a pronounced shrinking of the low-altitude projection of the plasma sheet region, reflecting the substantial reconfiguration of the magnetotail in RCM model during the substorm growth phase. Fig. \ref{fig3}i-\ref{fig3}l and Fig. \ref{fig3}m-\ref{fig3}p depict the magnetic field \(B_z\) distribution in the RCM equatorial plane and nightside ionosphere at different times, respectively. During the substorm growth phase, the stretching of magnetic field lines leads to a noticeable decrease in the \(B_z\) magnetic field over time. The \(B_z\) gradient from the inner magnetosphere to the RCM tail boundary is non-monotonic, with a distinct \(B_z\) hump at \(L > 20\), a feature also observed in reconstructed magnetic fields \cite<e.g.,>{Sitnov19:jgr,Sitnov21, Arnold22}. Further results from the RCM simulation for the 2-hour substorm growth phase in the RCM equatorial plane and ionosphere are available in Movie S1.\par

\begin{figure}[!htbp]
    \centering
    \includegraphics[width=0.5\linewidth]{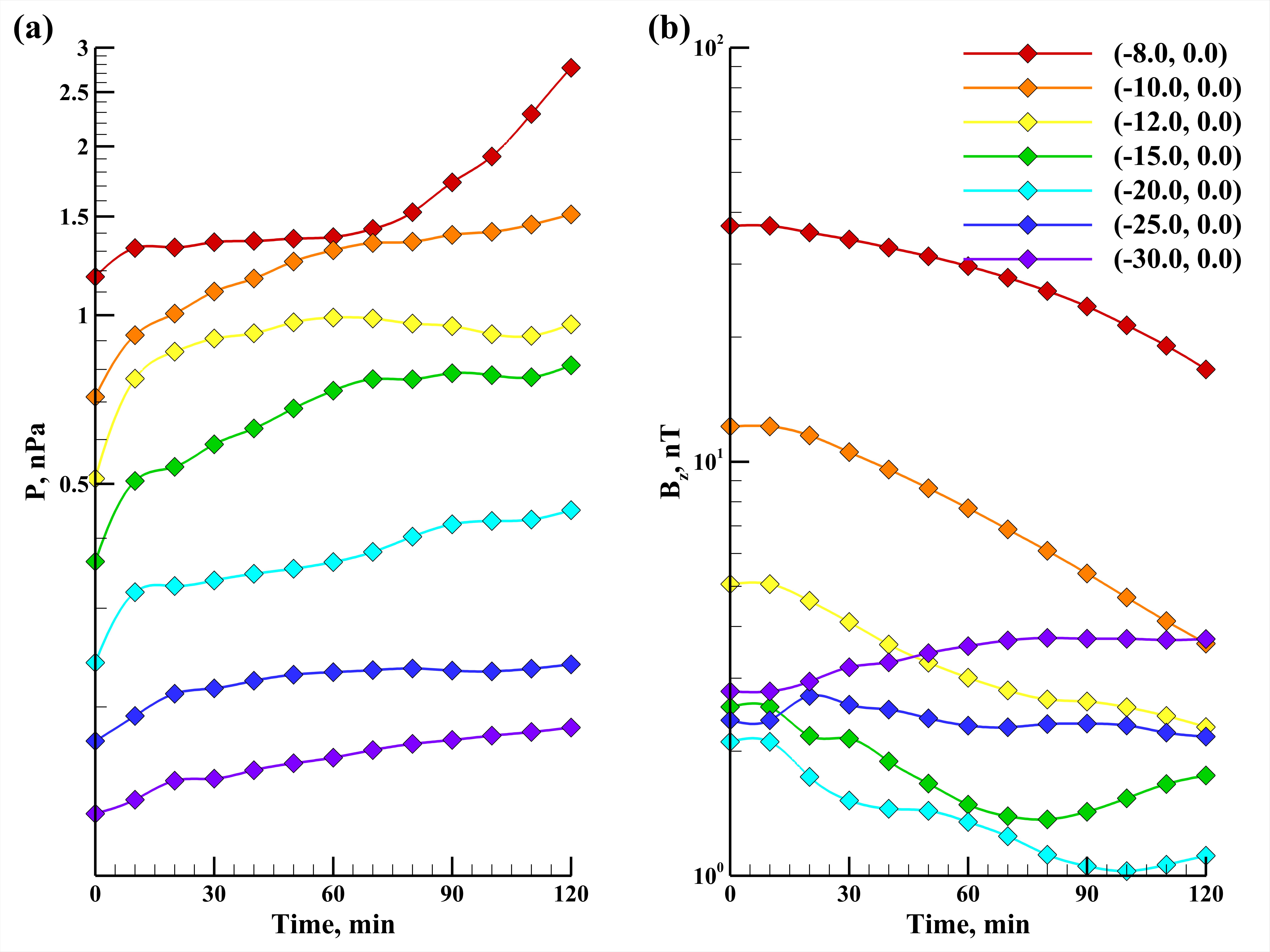}
    \caption{The variation of pressure and magnetic field \(B_z\) over time at different RCM equatorial locations marked by colored circles in Figure \ref{fig3}.}
    \label{fig4}
\end{figure}

Figure \ref{fig4} illustrates the temporal variations in pressure and magnetic field \(B_z\) at various equatorial locations, marked by the colored circles in Fig. \ref{fig3}a-\ref{fig3}d and \ref{fig3}g-\ref{fig3}i. Fig. \ref{fig4}a shows the evolution of pressure over time at these equatorial positions, revealing an increasing trend in pressure as time progresses. Notably, a high-pressure region, primarily composed of thermal protons, develops beyond the geosynchronous orbit after approximately 80 minutes, marking the interface between the outer radiation belt and the plasma sheet. Fig. \ref{fig4}b displays the temporal variation of the magnetic field \(B_z\) at the same locations, exhibiting a distinct \(B_z\) hump at \(L > 20\). This indicates the presence of a non-monotonic \(B_z\) gradient within the magnetotail.
This phenomenon likely reflects the complex structural changes occurring within the magnetosphere during the substorm growth phase \cite<see discussion in>{Hsieh&Otto14,Sergeev18:grl,Sitnov19:jgr,Shen23:jgr:ELFIN_dropout}. \par

\begin{figure}[!htbp]
    \centering
    \includegraphics[width=1\linewidth]{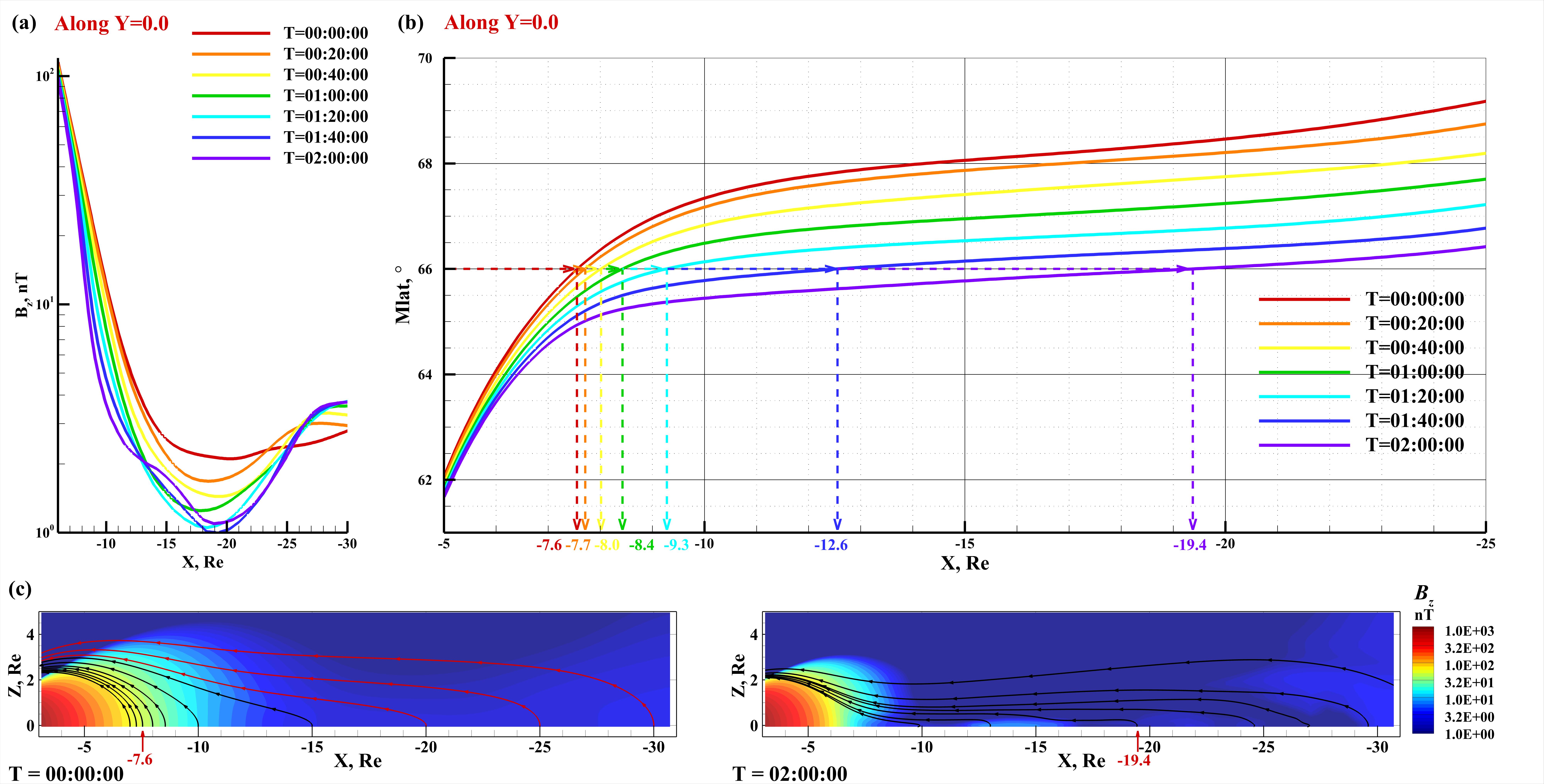}
    \caption{The RCM simulation results at different times along the radial direction at midnight. (a) Temporal variations of the radial profile of magnetic field \(B_z\) at midnight in the RCM equatorial plane. (b) Evolution of magnetic latitude over time for various radial positions at midnight. Dashed lines indicate the equatorial footprints of the magnetic field line at \(Mlat=66^\circ\), showing how it stretches to different \(L\) values at various times during the substorm growth phase. \blue{(c) Comparison of magnetic field configurations in the XZ plane at the beginning (\( T = \text{00:00:00} \)) and end (\( T = \text{02:00:00} \)) of the substorm growth phase. Background color contours show the \(B_z\) distribution. Ten magnetic field lines are traced from different equatorial positions at the initial time; among them, the seven field lines initially located closer to Earth (shown in black) predominantly remain within \(X < -30~R_E\) by the end of the simulation, as shown in the right panel. The remaining field lines (shown in red) extend farther tailward and are not displayed here. Red arrows highlight the specific magnetic field line discussed in panel (b).}}
    \label{fig5}
\end{figure}

Figure \ref{fig5}a offers further insights into the temporal variation of the radial profile of magnetic field \(B_z\) at midnight on the RCM equatorial plane. The distinct \(B_z\) hump region at \(L > 20\), as shown in Fig. \ref{fig4}b, appears even more pronounced here \cite<see discussion of observations of such humps in>{Sergeev18:grl,Sitnov19:jgr}. Fig. \ref{fig5}b shows the variation of magnetic latitude with $L$ at different times, highlighting the extent of magnetic field line stretching in the RCM results during the substorm growth phase. For instance, the magnetic field line footprint at \(Mlat=66^\circ\) is initially located at \(L=7.6\) in the equatorial plane at \( T = \text{00:00:00} \)
. Over time, this footprint stretches progressively: to \(L=7.7\) after 20 minutes, \(L=8.0\) after 40 minutes, \(L=8.4\) after 60 minutes, \(L=9.3\) after 80 minutes, \(L=12.6\) after 100 minutes, and finally to \(L=19.6\) by the end of the substorm growth phase simulation. This progression demonstrates that RCM successfully reproduces the gradual and significant stretching of magnetic field lines, a key feature of the magnetotail during the substorm growth phase. \blue{Figure \ref{fig5}c provides a more intuitive visualization of the stretching of magnetic field lines during the substorm growth phase. The two panels in Fig. \ref{fig5}c show the magnetic field \(B_z\) in the XZ plane at the initial time (\( T = \text{00:00:00} \)) and at the end of the growth phase (\( T = \text{02:00:00} \)), respectively. By comparing the two panels, it is evident that the plasma sheet region undergoes a significant reduction in \(B_z\) due to the stretching of magnetic field lines.  In the left panel, ten magnetic field lines with different equatorial plane footprints are traced at the initial time, with their footpoints distributed across a range of radial distances. Among them, the seven magnetic field lines initially located closer to Earth (shown as black solid lines) predominantly remain within \(X < -30~R_E\) at the end of the substorm growth phase, as shown in the right panel, while the remaining field lines (in red) extend farther tailward and are omitted for clarity. Additionally, the magnetic field line exemplified in Fig. \ref{fig5}b is highlighted with red arrows in Fig. \ref{fig5}c to facilitate comparison.} Overall, the RCM simulation effectively captures key characteristics of the substorm growth phase, including the reduction in magnetic field \(B_z\) and the progressive stretching of magnetic field lines, despite not accounting for all physical processes involved in the growth phase.\par

\section{ELFIN/RCM Comparison}\label{sec:comparison}

\blue{To validate our initial speculation that the variation in low-altitude electron fluxes is linked to magnetotail reconfiguration during the substorm growth phase \cite{Artemyev22:jgr:ELFIN&THEMIS},} we conducted the RCM simulation with a realistic setup. Specifically, we directly incorporated the energy-magnetic latitude spectra of trapped electrons observed by ELFIN to initialize the electron distribution within the corresponding energy and latitude ranges of the RCM prior to the current sheet thinning. For spatial domains or energy ranges not covered by ELFIN measurements, \blue{we utilized empirical models to initialize both thermal electron and proton distributions \cite{Lemon03, Tsyganenko&Mukai03}.} \blue{To account for the continuous supply of energetic electrons from Earth's magnetotail and to avoid depletion caused by dawn-dusk drift, we refreshed the ELFIN-derived initial electron distribution (derived from the trapped electron spectra) at every time step of the RCM simulation. This ensured that the evolution of electron fluxes at low altitudes was primarily driven by the effects of magnetic field reconfiguration during plasma sheet thinning, rather than by drift-related losses or other additional factors from the model affecting the electron distribution.} By maintaining consistency with the observed electron distribution throughout the simulation, this approach enables RCM to accurately capture the dynamics of electron fluxes during the substorm growth phase, aligning both with observations and theoretical predictions. \par

 Given that the RCM assumes a fully isotropic particle distribution, we cannot directly determine the location of the inner boundary of the plasma sheet from the ratio of \(J_{loss}\) to \(J_{trap}\), as is done in the ELFIN measurements. However, we can predict whether charged particles can be scattered nonadiabatically into the loss cone (when \(R_c/\rho < 8\)) or if the loss cone will remain depleted (when \(R_c/\rho > 8\)) based on the self-consistently computed magnetic field: the ratio of the magnetic field line curvature radius (\(R_c\)) to the particle gyroradius (\(\rho\)) controls the average magnitude of pitch-angle (PA) changes for charged particles \cite{Sergeev&Tsyganenko82,Delcourt94:scattering}. Along a particle's trajectory across the tail current sheet, the smallest \(R_c/\rho\) and the largest scattering occur at the equatorial point, where \(R_c \approx B_z / (\partial B_x / \partial z)\) and \(\rho = mc\sqrt{\gamma^2 - 1}/(eB_z)\). Here, \(c\) is the speed of light, \(m\) is the particle mass, \(e\) is the electric charge, and \(\gamma\) is the relativistic factor. For non-relativistic particles, \(\sqrt{\gamma^2 - 1} \approx \sqrt{2(\gamma - 1)} \approx \sqrt{2E / mc^2}\), where \(E\) is the particle's kinetic energy. If \(R_c/\rho < 8\), the pitch-angle (PA) scattering is strong, and the entire loss cone (LC) is filled isotropically during the traversal of the current sheet \cite<see discussions in>{Sergeev83,Wilkins23}. Conversely, if \(R_c/\rho > 8\), the scattering is weak, and the loss cone remains depleted. \par

\begin{figure}[!htbp]
    \centering
    \includegraphics[width=1\linewidth]{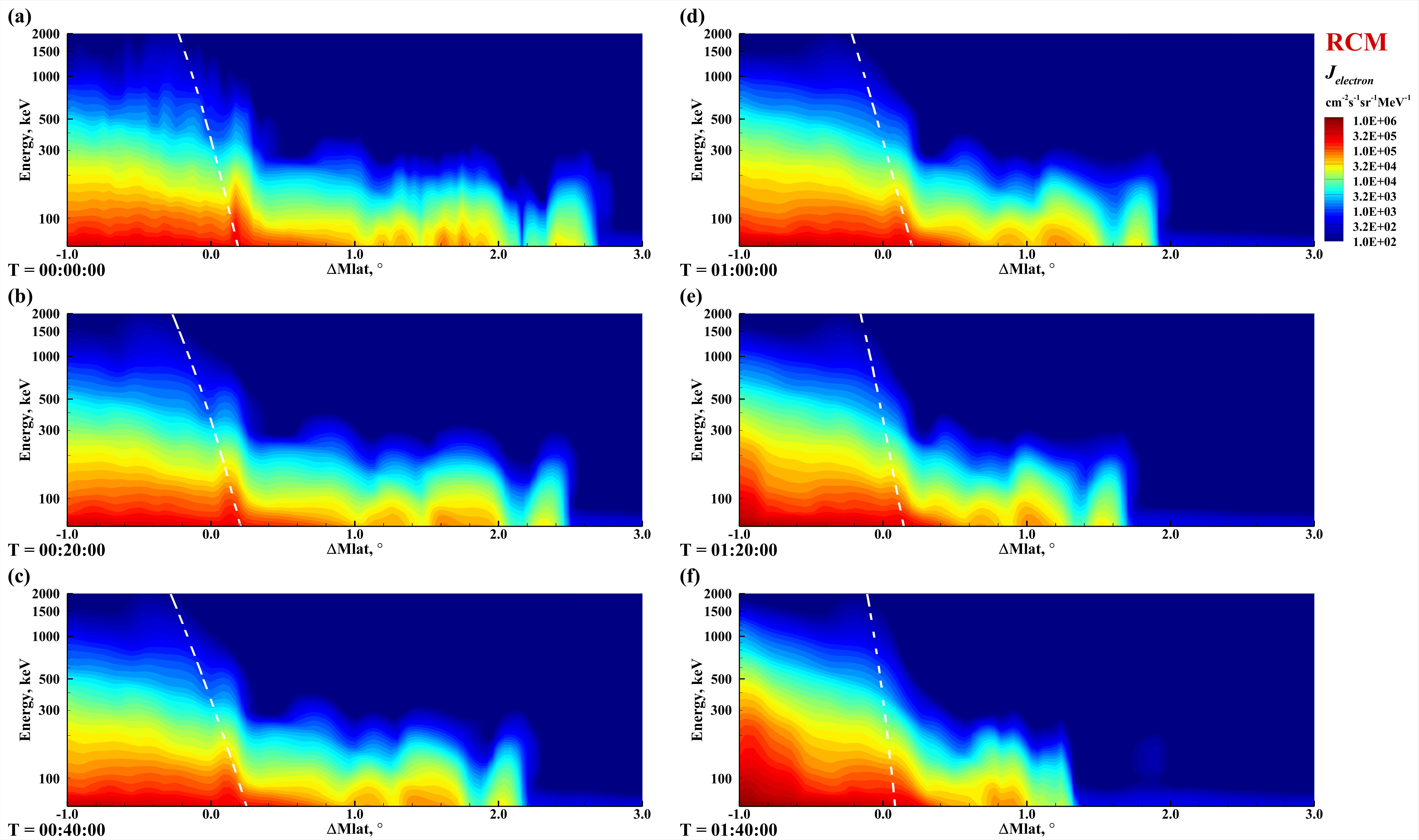}
    \caption{Overview of the RCM-simulated electron spectra for the ELFIN event in Fig. \ref{fig1} during the substorm growth phase. Each panel depicts the electron spectrum at a 20-minute time interval, with the white dashed line indicating \(R_c/\rho \sim 8\).}
    \label{fig6}
\end{figure}

Figure \ref{fig6} provides an overview of the RCM simulation results for the 100-minute substorm growth phase, with each panel representing a 20-minute time interval. As previously described, we directly used the trapped electron flux in Fig. \ref{fig1}a to initialize the electron distribution in the corresponding energy and latitude range in RCM at \( T = \text{00:00:00} \)
, as shown in Fig. \ref{fig6}a. Compared to Fig. \ref{fig1}a, our focus here is primarily on the plasma sheet region (\(\Delta Mlat > 0\)), because the radiation belt region (\(\Delta Mlat < 0\)) is not truly representative due to the absence of natural anisotropy. Moreover, we adjusted the initial magnetic field configuration of RCM to align the ELFIN-inferred plasma sheet inner boundary (where \(\Delta Mlat \sim 0\) in Fig. \ref{fig1}a-\ref{fig1}c) with the position calculated in the RCM (where \(R_c/\rho \sim 8\), which is energy-dependent and indicated by the white dashed line in Fig. \ref{fig6}a). Since the position of \(R_c/\rho \sim 8\) is energy-dependent, we aligned \(\Delta Mlat \sim 0\) with the latitude corresponding to the middle energy channel in the logarithmically spaced ELFIN channels, \blue{the middle of the white line.} Despite differences in the magnetic field configuration between the simulation and observations, which result in slight variations in the plasma sheet's latitude range (\(\Delta Mlat \sim 3.6^\circ\) in observations vs. \(\Delta Mlat \sim 2.6^\circ\) in the simulation), we ensured that the position of the plasma sheet inner boundary and the flux profile remain consistent between the two. As shown in the panels of Fig. \ref{fig6}, during the growth phase, the plasma sheet projection to low altitudes (seen as the latitudinal range of appreciable energetic electron fluxes) progressively shrinks over time, from \(\Delta Mlat \sim 2.6^\circ\) at \( T = \text{00:00:00} \) to \(\Delta Mlat \sim 1.3^\circ\) at \( T = \text{01:40:00} \). The gradient of the \(R_c/\rho \sim 8\) line becomes increasingly steep as time progresses. \blue{Due to the contraction of the plasma sheet projection region during the growth phase}, the flux levels comparable to those observed in the plasma sheet region before the growth phase can now only be measured within the increasingly narrow region tailward (polerward) of the \(R_c/\rho \sim 8\) line. For additional simulation moments and detailed changes in the electron spectrum, please refer to Movie S2. \blue{Although the 100-minute RCM simulation shows less contraction of the low-altitude projection of plasma sheet compared to the latitudinal extent inferred from the two ELFIN orbits separated by $\sim 3$ hours, it still accurately captures the main features of the low-altitude latitudinal shrinkage of the plasma sheet region associated with magnetotail reconfiguration.} Additionally, it is important to note that the continuous stretching of magnetic field lines throughout the RCM simulation domain leads to a corresponding reduction in the latitude range of the outer radiation belt (\(\Delta Mlat < 0\)).\par

\begin{figure}[!htbp]
    \centering
    \includegraphics[width=1\linewidth]{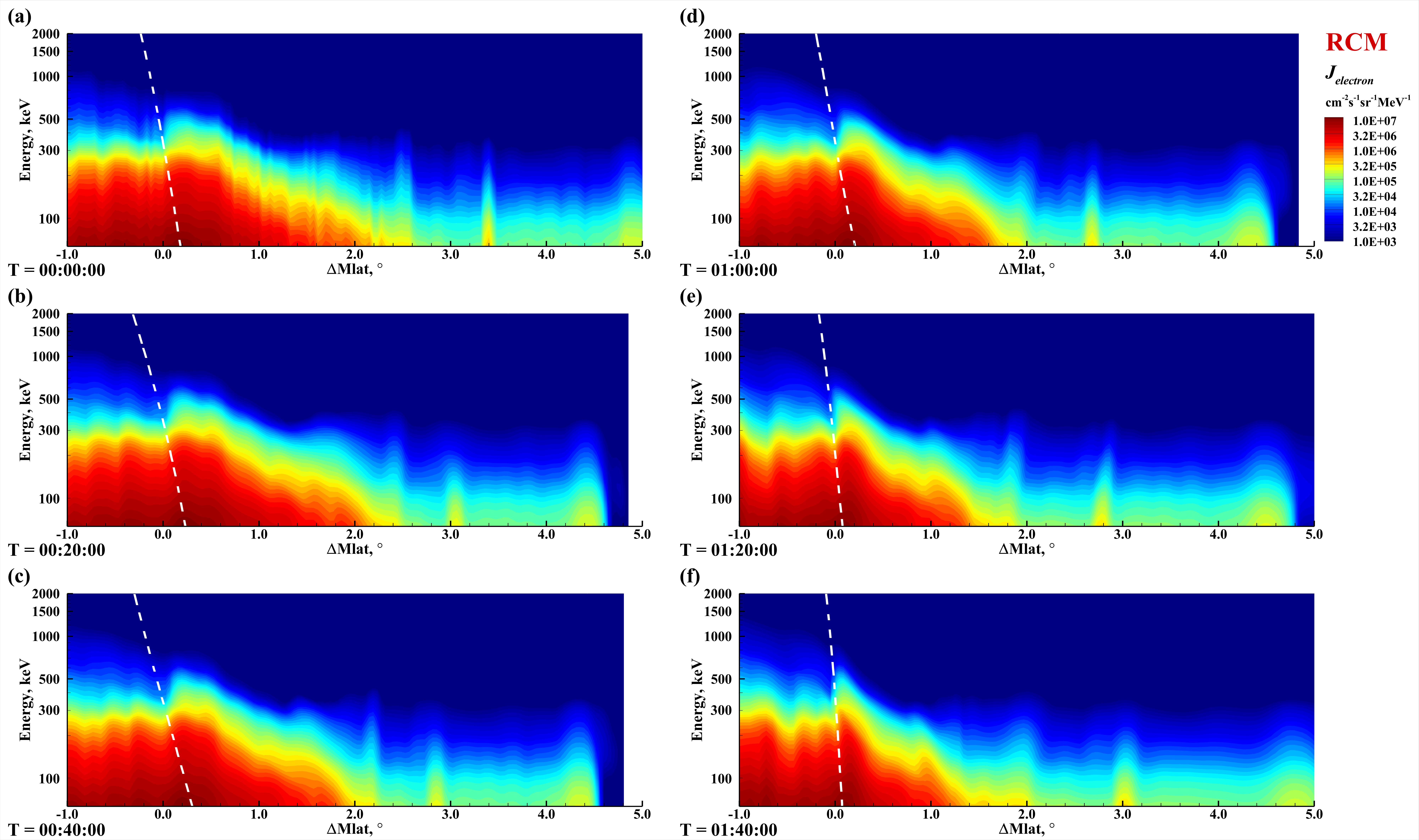}
    \caption{Overview of the RCM-simulated electron spectra for the ELFIN event in Fig. \ref{fig2} during the substorm growth phase, in the same format as Fig. \ref{fig6}.}
    \label{fig7}
\end{figure}

Figure \ref{fig7} presents the RCM simulation results for the ELFIN event shown in Fig. \ref{fig2}. The format mirrors that of Fig. \ref{fig6}, illustrating the 100-minute substorm growth phase, with each panel representing a 20-minute interval. Since we directly use the trapped electron flux from Fig. \ref{fig2}a to initialize the energetic electron distribution in RCM, the resulting spectra at \( T = \text{00:00:00} \) (Fig. \ref{fig7}a) closely resemble Fig. \ref{fig2}a in the plasma sheet region (\(\Delta Mlat > 0\)). Similarly, although the absolute latitude range of the plasma sheet (\(\Delta Mlat > 0\)) differs between the simulation and observations, the position of the plasma sheet inner boundary and the flux profile are kept consistent. As shown in Fig. \ref{fig7}, the plasma sheet region with relatively high electron fluxes (primarily represented in red) gradually shrinks during the growth phase, contracting from around \(\Delta Mlat < 2.3^\circ\) at \( T = \text{00:00:00} \) to \(\Delta Mlat < 1.4^\circ\) at \( T = \text{01:40:00} \). However, the low electron flux tail at higher latitudes (visible at \(\Delta Mlat > 4.0^\circ\) in Fig. 2a and \(\Delta Mlat > 2.3^\circ\) in Fig. 7a) shows minimal shrinkage, likely due to boundary effects near the RCM's limits. For further insights into the simulation results and a more detailed analysis of the electron spectrum changes, please refer to Movie S3. Although the shrinkage of the main high-flux plasma sheet region is less pronounced than in Fig. \ref{fig6}, likely due to differences in the magnetic field configurations between the two events, the RCM simulation still effectively captures the key features of the plasma sheet's latitudinal contraction.

\section{Discussion and Conclusions}\label{sec:discussion}
This study tests the hypothesis that the shrinking of the low-altitude projection of the plasma sheet (characterized by isotropic energetic electron fluxes within $[50,300]$keV) can be attributed to magnetic field reconfiguration in the magnetotail during \blue{the current sheet thinning (presumably within the substorm growth phase).} This hypothesis is important for advancing low-altitude monitoring of magnetotail dynamics, a promising approach for revealing the characteristics of pre-onset magnetic field configurations and the dynamics of plasma injections during the expansion phase \cite<see discussion in>{Sergeev18:grl,Millan&Ukhorskiy24,Artemyev24:jgr:ELFIN&IBe}. Two potential approaches can be used to test this hypothesis about the fine structure of low-altitude electron fluxes in the plasma sheet projection \cite<see, e.g.,>{Shen23:jgr:ELFIN_dropout,Shen23:jgr:ELFIN_THEMIS}: comparing low-altitude energetic flux measurements with advanced empirical magnetic field models \cite<e.g.,>{Kubyshkina11,Kubyshkina15,Sergeev12:IB,Shi&Stephens24:arxiv} or using simulations of the magnetotail. In this study, we adopted the latter approach and used RCM simulations, which have demonstrated strong capabilities in reproducing substorm magnetotail features \cite{Yang11,Yang13,Sergeev21:rcm,Sergeev23:rcm}. Note that alternative simulations, such as global MHD models \cite{Gordeev17,Eshetu18,Sorathia18}, could also be adopted in such comparisons.

By combining RCM simulations of substorm magnetic field reconfiguration with energetic electron fluxes observed by ELFIN before the substorm to initialize the flux distribution, we simulated the evolution of the plasma sheet's energetic component under the model's electric and magnetic fields during the substorm growth phase. We then compared the low-altitude projection of this evolution to subsequent ELFIN observations. Although the RCM assumes isotropic fluxes and cannot explicitly determine the isotropy boundary demarcating the plasma sheet and outer radiation belt, we used the RCM magnetic field model to estimate this boundary based on equatorial conditions of electron scattering. \blue{The simulation results demonstrate the formation and evolution of a narrow band of energetic electron fluxes poleward of the isotropy boundary (approximated by the \(R_c/\rho = 8\) threshold),  whose latitudinal confinement and morphology closely resembles the low-altitude ELFIN observations \cite<see more examples in>{Artemyev22:jgr:ELFIN&THEMIS}.} These results lead to two main conclusions:
\begin{itemize}
  \item The substorm dynamics of the latitudinal range of low-altitude projections of the plasma sheet's energetic population are driven by magnetic field line reconfiguration in the magnetotail. Thus, this phenomenon provides a valuable means of tracing magnetic field configurations when combined with sufficiently advanced magnetic field models. This conclusion builds on ideas previously proposed and tested in \cite{Sergeev93:test,Kubyshkina11}, which involved modifications to the \cite{Tsyganenko89,Tsyganenko95} magnetic field models using low-altitude POES observations. Recent advancements in empirical magnetic field reconstruction techniques \cite{Stephens19, Sitnov19:jgr}, combined with ELFIN substorm observations \cite{Shi&Stephens24:arxiv}, have further demonstrated the growing potential of this approach. Our results suggest that simulations like the RCM can also benefit from {\it calibration} with low-altitude measurements of energetic particle fluxes.
  \item An important element of the RCM model is the system of field-aligned currents that form during substorm dynamics, coupling the magnetosphere and ionosphere current systems \cite{Yang12,Yang14,Wei22}. The location of these field-aligned currents relative to the inner edge of the plasma sheet (or magnetotail current sheet) is particularly important for addressing the challenges of ionosphere-magnetosphere magnetic mapping \cite<see discussion in>{Sergeev12:IB, Sergeev20:aurora, Jiang12:fast&asi_aurora, Liang13:IB}. Our results demonstrate the strong potential for further integration of field-aligned current simulations with low-altitude observations of the plasma sheet, offering new opportunities for advancing our understanding of ionosphere-magnetosphere coupling.
\end{itemize}
\blue{Therefore, the main conclusion of this study is the confirmation that the latitudinal shrinking of the low-altitude projection of the plasma sheet (energetic electron component) can be explained by the magnetic field line reconfiguration during magnetotail current sheet thinning. This confirmation opens two directions for the investigation of magnetotail dynamics using low-altitude observations. First, the new generation of simulations and models of magnetic field configurations \cite{Sitnov19:jgr,Stephens19,Andreeva&Tsyganenko19} may incorporate such low-altitude measurements to improve tracing of  substorm dynamics \cite<see>{Shi&Stephens24:arxiv}. Second, low-altitude dynamics of the energetic electron component of the plasma sheet can be compared with dynamics of field-aligned current systems, which are also strongly associated with magnetic field reconfiguration \cite<see discussion in>{Yahnin97,Sergeev12:IB}. Such a comparison will further advance our modeling of these two elements of magnetosphere-ionosphere coupling -- field-aligned currents carried by low-energy electrons and the precipitation of energetic electrons scattered within the magnetotail current sheet.}

\acknowledgments
W.S., X.J.Z., and A.V.A. acknowledge support by NASA awards 80NSSC23K0100, 80NSSC23K0108 and NSF award 2400336.

We are grateful to NASA’s CubeSat Launch Initiative for ELFIN's successful launch in the desired orbits. We acknowledge early support of ELFIN project by the AFOSR, under its University Nanosat Program, UNP-8 project, contract FA9453-12-D-0285, and by the California Space Grant program. We acknowledge critical contributions of numerous volunteer ELFIN team student members.

\section*{Open Research}
\noindent Electron fluxes measured by ELFIN are available in CDF format \cite{elfin_data}.  THEMIS magnetic field, plasma moments, and electron fluxes are available at \cite{themis_data}. Data analysis was done using SPEDAS V4.1 \cite{Angelopoulos19}. The software can be downloaded from {http://spedas.org/wiki/index.php?title=Downloads\_and\_Installation}. The RCM simulation output can be accessed through \cite{weiqin_data}.



\end{document}